# A Novel Skiagraphic Method of Casting Shade of a Torus

Tanvir Morshed[a]*

[a]Department of Architecture, Stamford University Bangladesh, Dhaka, Bangladesh; *corresponding author

**Abstract:**

This paper introduces a novel skiagraphic method for shading toroidal forms in architectural illustrations, addressing the challenges of traditional techniques. Skiagraphy projects 3D objects onto 2D surfaces to display geometric properties. Traditional shading of tori involves extensive manual calculations and multiple projections, leading to high complexity and inaccuracies. The proposed method simplifies this by focusing on the elevation view, eliminating the need for multiple projections and complex math. Utilizing descriptive geometry, it reduces labor and complexity. Accuracy was validated through comparisons with SketchUp-generated shading and various torus configurations. This technique streamlines shading toroidal shapes while maintaining the artistic value of traditional illustration. Additionally, it has potential applications in 3D model generation from architectural shade casts, contributing to the evolving field of architectural visualization and representation.

## 1. Introduction

The practice of examining and rendering shades and shadows, known as skiagraphy, sciography, or shadowgraphy, is widely utilized in architectural illustrations. It falls within the realm of descriptive geometry, a discipline focused on projecting three-dimensional objects onto a two-dimensional surface to uncover their geometric characteristics and relationships. It revolves around the principles of creating projections of forms to acquire insights into the geometric properties and its relationship with surrounding forms. Conversely, skiagraphy offers valuable insights into the contours, shapes, and environmental aspects[1] of architectural designs with the help of shades and shadows cast. The skiagraphic or manual process of capturing shades and shadows is considered quite challenging due to intricate geometries, limited reference materials, and the growing popularity of computer-generated shading.

---

[1] Shades and shadows in a drawing provide valuable insights into the environmental aspects of a design, indicating how it would perform under various elements. Additionally, they contribute to understanding the aesthetic aspects of the design.

Over the past two centuries, architectural illustrations have often placed significant emphasis on the precise calculation of shades and shadows, a process heavily reliant on manual calculations of geometric interconnections of forms. While simpler forms are generally more manageable, ornate classical decorations with intricate details have remained popular among designers, despite the inherent challenges of casting shadows for these multifaceted shapes. Complex forms such as floral designs, toroids, cones, spheres, hollow shapes, and combinations of primitives pose substantial difficulties in the shading process.

Traditional architectural illustration, particularly in depicting shades and shadows of intricate elements, requires a combination of mathematical precision and artistic skill. Skiagraphy, the art of shadow projection, relies on descriptive geometry and complex formulas. The conventional method involves creating plan and elevation views on a single sheet, using numerous construction lines to plot shadow fall across complex geometries like Corinthian capitals or baroque ornamentation. This process often requires multiple drawing overlays to build up the final shadow projection. It's time-consuming and meticulous, involving extensive erasing of temporary lines to achieve the final, refined drawing. The method is particularly challenging for complex classical forms with their abundance of intricate details. This laborious method underscores the high level of skill and patience demanded of architectural illustrators in the pre-digital era. While modern computer-aided design has revolutionized this process, understanding these traditional techniques provides valuable insights into the historical development of architectural representation and the enduring challenge of accurately depicting light and shadow in complex architectural forms.

In fact, from the 1960s onwards, architectural and engineering drawings, as well as skiagraphy, played a significant role in assessing shades and shadows within computer-generated 3D environments. Various shading techniques, including but not limited to flat shading, Gouraud shading, and Phong shading, were developed to improve the depth, color transitions, and precision of shading and highlights in computer graphics. In the 1970s and early 1980s, several computational methods such as Shadow Mapping and Ray Tracing emerged. Techniques like Radiosity and Ambient Occlusion were developed in the 1990s, followed by Voxel Cone Rendering in the 2010s. From the 1980s onwards, the development of Radiosity, Path Tracing, Photon Mapping, Ambient Occlusion, and Precomputed Radiance Transfer (PRT) shaped the evolution of global illumination and real-time ray tracing techniques, which are integral to modern gaming and architectural renderings (Appel, 1968) (Goral et al., 1984) (Jensen, 1996) (Phong, 1998) (Talbot, 1999) (Bunnell, 2005) (Crassin et al., 2011) (Sloan et al., 2023). These advancements greatly enhanced the representation of shades and shadows in computer graphics. It is important to recognize that these advancements are indebted to a history spanning over two and a half millennia, drawing from various fields, including mathematics, optics, construction technology, and its documentation, as well as developments in art, architecture, and architectural illustrations. In particular, perspective drawing techniques, projection drawings, and descriptive geometry have played pivotal roles in shaping these innovations.

This paper contributes to the ongoing evolution of shade-casting techniques in architectural illustration by introducing a novel graphical approach for shading a torus. The proposed method focuses primarily on shading in the elevation view, specifically on the vertical plane. It demonstrates that projecting the torus solely in elevation is sufficient for accurate shade calculation in that view. A key advantage of this approach is the elimination of the need to draw both plan and elevation views, whether on a single sheet or separate sheets. This simplification significantly reduces the complexity and labor traditionally associated with shade projection for complex forms.

While acknowledging advancements in computer-generated graphics, this method aims to preserve and simplify the practice of hand-drawn architectural illustrations. By exploring more efficient manual techniques, this approach seeks to maintain the artistry and craft of traditional architectural representation in the digital age. This streamlined method not only offers practical benefits for illustrators but also contributes to the broader goal of balancing traditional skills with modern technological advancements in architectural visualization.

## 2. Skiagraphy or Sciography

Skiagraphy, from Greek "skia" (shadow) and "graphein" (to write), is a branch of perspective in architectural drawing focusing on shadow projection and object representation under varying light conditions. In architecture, it examines how forms cast shadows on flat surfaces or view planes. This discipline requires a deep understanding of light-object interactions and shadow formation, bridging artistic representation and scientific observation. Skiagraphy is crucial in architectural drawing for creating depth, dimension, and realism, effectively communicating design intent and spatial qualities.

However, in descriptive geometry, shades and shadows are symbolic representations of light's impact on three-dimensional objects when projected onto a two-dimensional plane. These techniques are crucial for visualizing and understanding spatial relationships and object shapes. "Shades" refer to darker areas on an object that are partially or fully shadowed, shielded from direct light. These typically occur on surfaces facing away from the light source or obscured by other parts of the object. Shades are often depicted through uniform darkening or hatching of relevant sections. "Shadows," in contrast, are darkened areas cast by an object blocking light. These appear on surfaces not directly facing the light source or on other objects in the scene. Both shades and shadows play a vital role in conveying depth, form, and spatial relationships in architectural and engineering drawings. Their accurate representation enhances the overall comprehension of three-dimensional forms in two-dimensional representations, making them essential tools in visual communication for these fields. This approach to representing light and shadow effects bridges artistic representation with technical accuracy, providing a powerful means of communicating complex spatial information in a visually intuitive manner.

Skiagraphy is essential in architectural design, allowing architects to analyze and manipulate light's interaction with structures. By examining shade and shadow patterns, designers can create visually compelling spaces and optimize natural light in their designs. This technique bridges the gap between two-dimensional plans and three-dimensional realities, aiding in the visualization of spatial relationships and forms. It helps predict how designs will interact with light throughout the day and seasons, guiding decisions on orientation, form, and fenestration. Additionally, skiagraphy enhances communication of design intent, conveying the experiential qualities of spaces to clients and collaborators, and influences both aesthetic and functional aspects of architecture. It remains an indispensable skill, blending artistic vision with technical precision to create buildings responsive to light and shadow.

## 2.1. Architectural Skiagraphy or Sciography Convention

The skiagraphy convention, employed for casting shadows and shades, imparts a sense of depth to architectural structures through the projection of light rays (Lr) at a 45° angle in both plan and elevation projections. The true angle ($A^t$) of these rays is precisely set at 35°16', as determined by the rotation of Lr in parallel alignment with either the horizontal (H) or vertical (V) coordinate planes. In Figure 1e, $A^t$ is the genuine angle, formed by revolving Lr in parallel with the vertical plane, thus ensuring accurate shadow representation. This methodology results in shadows that mirror the width or depth of the objects upon which the shadows are cast (refer to Figure 1).

This scientific approach, vital for creating mechanical representations of shades and shadows, also possesses an artistic dimension because it favors manual, hand-drawn techniques. Unlike complex construction blueprints, these drawings enhance spatial visualization for both professionals and non-architects. They show how shapes relate, reveal depth, and highlight the beauty of proportions, making designs more understandable, even for non-architects. For instance, the absence of shades and shadows in a projection-based visual environment, as shown in Figure 2, leads to a lack of realism. These elements are, in essence, considered the "fourth dimension", capturing the temporal aspect of 3D geometry, much like in real-life scenarios.

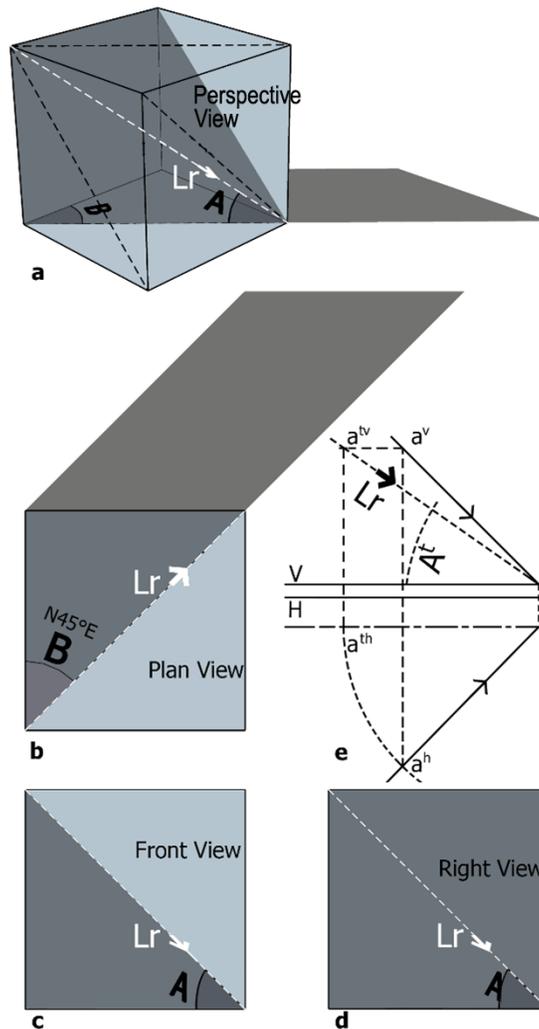

*Figure 1: Skiagraphy Convention in Architecture or Architectural Shade and Shadow Convention a. Perspective, b. Plan View (Top-Down Projection), c. Front View (Frontal Elevation), d. Right View (Side Elevation), and e. Graphical Representation of Shadow Angles.*

## 2.2. Traditions of Torus Skiagraphy

The term "torus" finds its origins in Latin, signifying a swelling or bulging shape. It represents a surface or solid created by rotating a closed curve, such as a circle, around an axis of rotation lying in the same plane but not intersecting it. This geometric shape is akin to a ring doughnut. In the realm of architecture, a torus denotes a substantial convex molding, typically semicircular in the cross-section, usually positioned at the base of a column. It holds significant importance in various classical geometric depictions of architecture which are practiced even today.

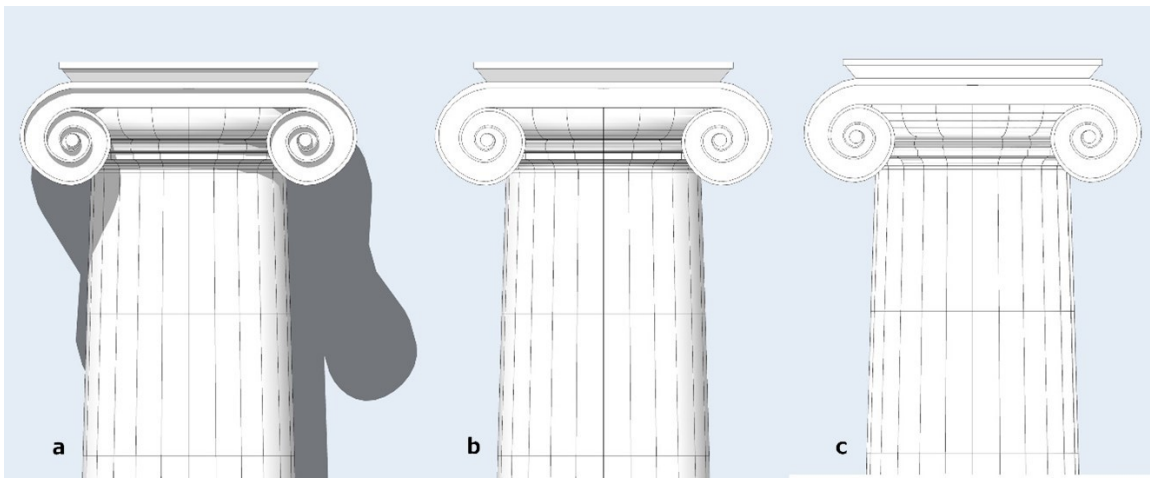

*Figure 2: Impact of Lighting on Spatial Perception a. Full Shadow Rendering, b. Shading Without Cast Shadows, and c. Unshaded Geometric Outlines.*

Architects, designers, and mathematicians frequently use mathematical methods to add depth to architectural drawings through shading and shadowing. Precise representation of light and shadow on structures enhances their aesthetic appeal. However, existing literature for rendering shades and shadows on toroidal forms can be complex and time-consuming.

Table 1 provides a summary of common techniques for creating shades and shadows in literature, with a focus on their complexity. Each method is given a score between 1 and 5, where 1 is straightforward and 5 is highly intricate. This score is determined by evaluating factors such as the number of stages required, the production of various drawings (including plans, elevations, and sections), and angle calculations. Drawing a plan and an elevation together accounts for one point, while additional drawings like sections and projections each contribute another point. Methods requiring more than 30 steps are automatically assigned the highest complexity score. Angular calculations contribute one additional point to the complexity score, capped at five angles. Methods requiring more than five angle measurements don't receive further points. Interestingly, the most complex methods often involve unique calculations or geometric representations. Despite their complexity, these methods are explored for their novel approaches to shade and shadow computation, although they often prove inaccurate when applied to various toroidal shapes.

Table 1: Selected Prominent Torus Shadow Projection Techniques

| Sl. No. | Author/Book/Document Name | Fig. or Plate no. | Multiple Projection | Number of steps | Complexity metrics | Reference |
|---|---|---|---|---|---|---|
| 1 | F. N. Wilson | Fig. 381 | Yes | 30 | 4 | (Willson, 1898) |
| 2 | Wooster Bard Field and Thomas E. French | Plate 14 | Do | 26 | 5 | (Field and French, 1922) |
| 3 | Henry McGoodwin | Fig. 43 | Do | 26 | 4 | (McGoodwin, 1904) |
| 4 | William R. Ware | Fig. 110 | Do | 25 | 5 | (Ware, 1913) |
| 5 | William R. Ware | Fig. 111 | Do | 25 | 5 | (Ware, 1913) |
| 6 | William R. Ware* | Fig. 112 | Do | 16 | 3 | (Ware, 1913) |
| 7 | William R. Ware | Fig. 113 | No | 22 | 5 | (Ware, 1913) |
| 8 | William R. Ware | Fig. 115 | Yes | 40 | 5 | (Ware, 1913) |
| 9 | Joseph Gwilt | Fig. 7 | Do | 28 | 5 | (Gwilt, 1824) |
| 10 | J. J. Pillet ** | Fig. 47 | Do | 35 | 5 | (Pillet, 1896) |
| 11 | Par Charles Normand | Fig. 5, Plate 1 | Do | 26 | 5 | (Normand, 1838) |
| 12 | Par Charles Normand | Fig. 3, Plate 4 | Yes | 38 | 5 | (Normand, 1838) |
| 13 | Cosimo Rossi Melocchi | Fig. 12 | Do | 35 | 5 | (Melocchi and Tofani, 1805) |
| 14 | David Christoph Lange | Fig. 52 | Do | 23 | 4 | (Lange, 1921) |
| 15 | Jullian Millard*** | Fig. 47 | Do | 35 | 5 | (Pillet, 2018) |
| 16 | Harry W. Gardner | Fig. 52 | Do | 22 | 3 | (Gardner, 1905) |
| 17 | John H A McIntyre | Fig. XIII | Do | 55 | 5 | (M'Intyre, 1901) |
| 18 | The University of Illinois Shadow Notes | Fig. 43 | Do | 20 | 3 | (Illinois (Urbana-Champaign), 2021) |

* Chosen for simplicity, but results vary across different torus shapes.

** This technique employs mathematical formulas for shadow computation, contrasting with intuitive, non-formulaic approaches.

*** Translation of J J Pillet's book by Julian Millard. See sl. no. 10 in the table above.

Given the complexity of existing torus shadow projection methods, this paper introduces a new, streamlined approach in Section 3. This user-friendly method caters to architects and mathematicians who prefer a descriptive geometric approach, eliminating the need for complex formulas and intricate drawing references. Furthermore, this proposed method offers improved accuracy compared to existing techniques, making it a valuable tool for architectural design and illustration.

## 3. Proposed Torus Skiagraphy

Following an in-depth analysis of the intricate nature of torus skiagraphies, as evident in the table above, a groundbreaking method for casting torus shades is introduced here (Fig. 3). This approach focuses solely on the vertical plane, or the elevation/side view, for shade casting, while shadow casting has not been explored due to the presence of straightforward methods in existing literature. It's designed for accessibility, allowing architects, designers, and mathematicians to sketch without resorting to complex mathematical equations. It adheres rigorously to the fundamental principles of descriptive geometry, making the process straightforward and devoid of any requirement for referencing architectural plans or sections. The resulting torus shades are then thoroughly examined in the subsequent section, showcasing a diverse range, from ring toruses to distinctive horn shapes (Figs. 4, 5, and 6). This comprehensive analysis encompasses a wide array of shade formations, spanning bulky and slender, as well as wide and narrow torus configurations.

In Figure 3, the central line AB is drawn through the centers of the torus's rings in the vertical plane. Points D and E mark the lowest and highest meridian traces of the shade line, respectively. These points are determined by the intersections of 45° lines drawn from the ring centers. Line DE bisects AB at point C. The ring centers create points P and Q on AB. Points H and I originate from perpendicular line drawn from D and E, respectively, and meet at point C when connected. Lines HP and IQ intersect DE at points J and K, respectively. Line LM runs through point C, remaining parallel to HP and IQ. Points L and M represent the extensions of lines perpendicular to AB from points J and K, respectively. The extensions of HD and IE form points n and o at An and Bo, respectively. Lines Pn' and Qo' are the mirror images of Pn and Qo with respect to PD and QE, respectively. Point n' is derived easily by drawing a perpendicular from D to the extension of the torus's outer edge line and so is for o'. Similarly, lines Pn" and Qo" mirror Pn' and Qo' with respect to lines parallel to HI passing through points P and Q, respectively. Lines Pn" and Qo" intersect the rings at points n" and o", respectively. Tracing the points P, D, n", H, M, Q, E, o", I, L, and back to P results in the creation of the shade path of the toroidal shape.

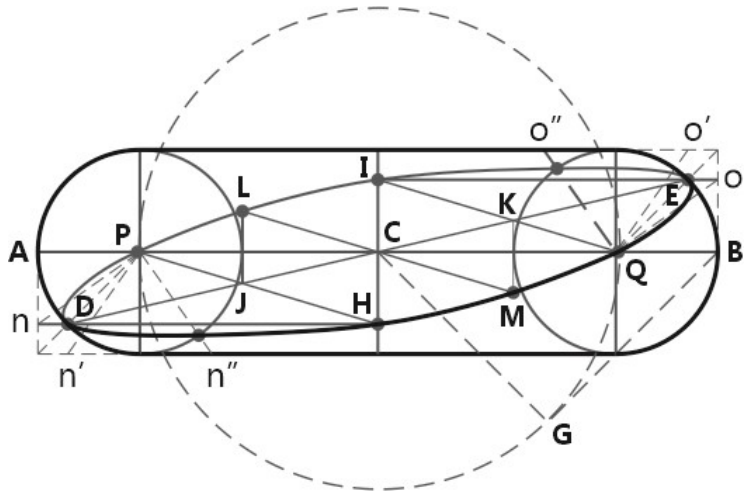

*Figure 3: The Proposed Method for Determining the Shade of a Torus*

This method, as outlined in Section 2.2, has a complexity score of 2, making it the simplest among all the approaches summarized in Table 1 above. This is a significant advantage, as it suggests that this method is relatively straightforward to implement and understand. Furthermore, the analysis offered in the next section is grounded in the practical context of a classical toroidal column base. This is an important consideration, as it indicates that the method has been evaluated in a real-world scenario, rather than being purely theoretical. The use of a classical toroidal column base provides a concrete application for the method, allowing the author to assess its performance and suitability in a tangible, practical setting. This grounds the analysis in a relatable and relevant context, making the findings more meaningful and applicable to readers who may be interested in using this method in similar architectural or engineering projects. By situating the analysis within the context of a classical toroidal column base, this methos demonstrates the method's potential for practical implementation and its ability to address real-world design challenges. This contextual grounding adds depth and relevance to the discussion, making the overall presentation more compelling and useful for readers.

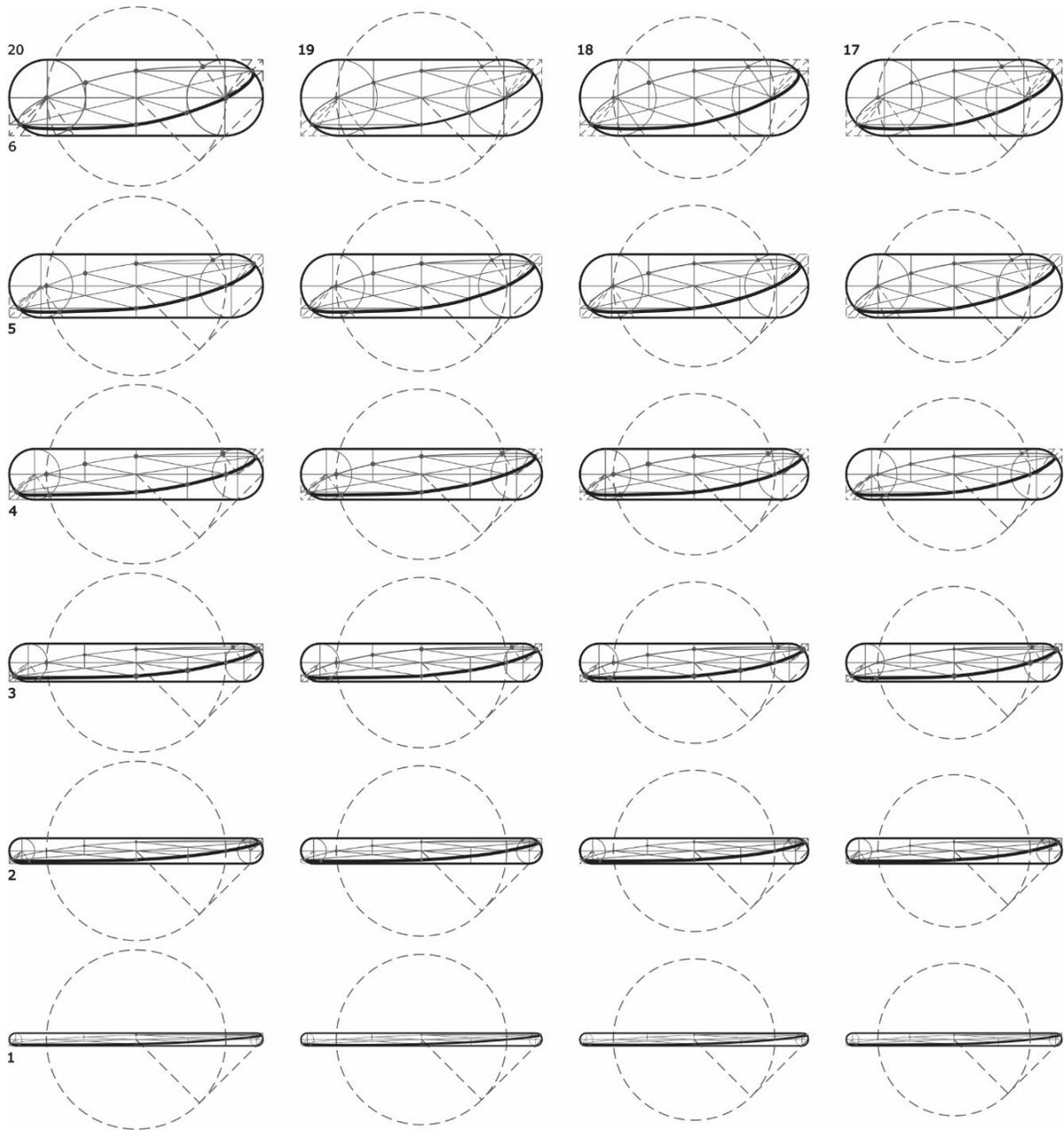

*Figure 4: This figure presents 24 scaled-down representations of toroidal shade casts. The original CAD drawings ranged in size from 20 to 17 inches wide (left-to-right) and 6 to 1 inch tall (top-to-bottom). Here, the depictions are adjusted to fit the standard dimensions of a journal publication.*

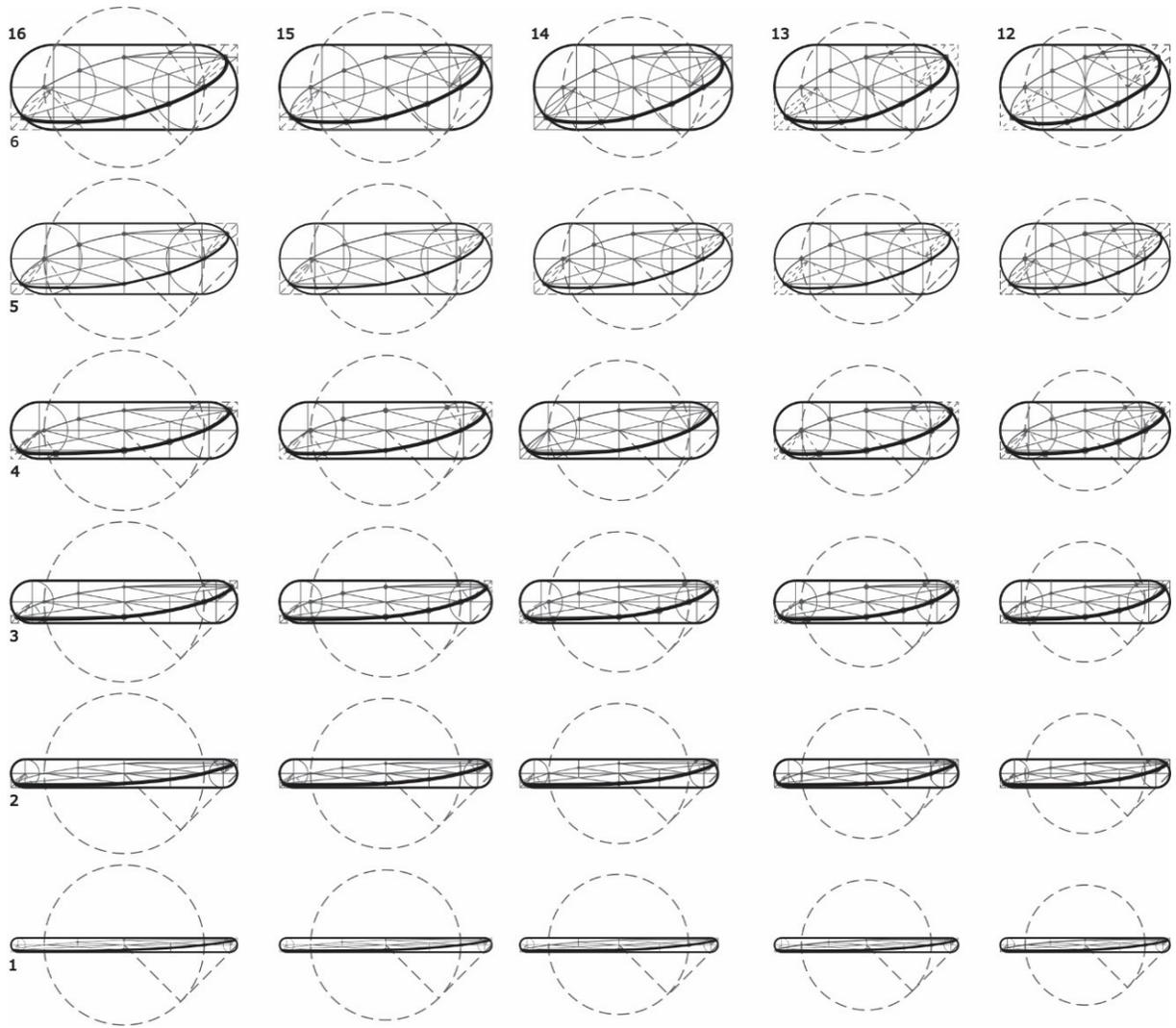

*Figure 5: 30 Scaled-Down Versions of Torus Shade Cast Generated from 16 to 12 Inches (width) From Right to Left and 6 to 1 Inch (Height) From Top to Bottom. This Depiction Scaled to Fit in Papers of Typical Journal.*

## 3.1. Analysis of the Proposed Methods

The matrix presented in Figures 4, 5 and 6 displays the shaded representations of toruses, serving as evidence of the proposed method's effectiveness in capturing shade of toroidal surface in various configurations. These scaled-down versions of the toruses, along with their respective cast shades, have been adjusted to accommodate the typical journal paper dimensions. The original computer-aided design (CAD) drawings, however, ranging in width from 20 to 12 inches in the left-to-right direction and from 6 to 1 inch in the top-to-bottom direction. This diversity in sizes ensures that the analysis of shades validates the proposed method's accuracy across a broad spectrum of torus shapes.

The matrix presented in Figures 4, 5, and 6 displays the shaded representations of toruses, serving as evidence of the proposed method's effectiveness in capturing the shade of the toroidal surface in various configurations. These scaled-down versions of the toruses, along with their respective cast shades, have been adjusted to accommodate the typical journal paper dimensions. This adjustment ensures that the visual representations are appropriately sized for the publication format. However, the original computer-aided design (CAD) drawings range in width from 20 to 12 inches in the left-to-right direction and from 6 to 1 inch in the top-to-bottom direction. This diversity in sizes (ranging from 20 to 12 inches wide and 6 to 1 inch tall) is a crucial aspect of the analysis, as it ensures that the evaluation of the shades validates the proposed method's accuracy across a broad spectrum of torus shapes.

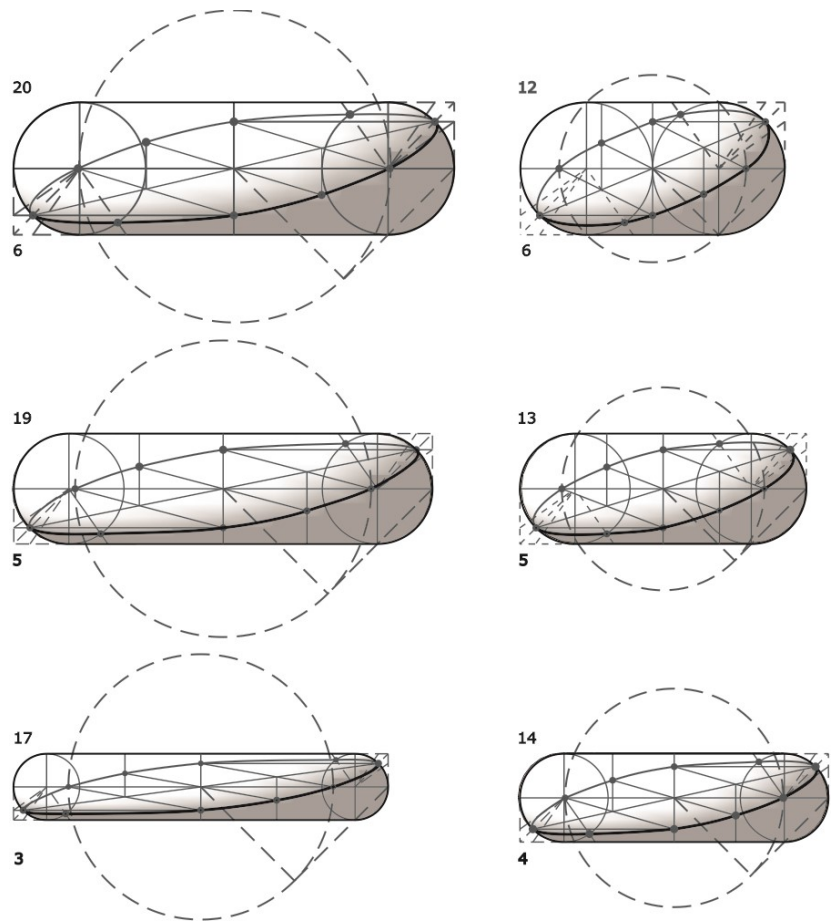

*Figure 6: This Figure Presents a Comparison Between the Shading Generated by the Proposed Method and the Shade Casts Created by SketchUp Software. The Depictions are Adjusted to Fit the Standard Dimensions of a Journal Publication.*

The results were then compared to the shading produced by SketchUp software, revealing a strong agreement between the outlines generated by this method and those created by SketchUp. To illustrate these correspondences more clearly, Figure 6 presents a selection of randomly chosen examples showcasing the alignment between the shade outlines from both CAD and SketchUp.

## 4. Conclusion

Skiagraphy, a widely embraced technique in architectural design and illustration, leverages the use of shadows to impart depth perception to architectural creations. Shade and shadow casting of classical ornamentation, characterized by its intricacy, enjoys a substantial following within the design community. However, conventional skiagraphy methods, particularly when dealing with complex shapes like tori, can pose significant challenges.

This paper introduces an innovative approach to casting shades on toroidal forms, offering a simplified and more user-friendly alternative to traditional techniques. This straightforward method employs graphical tools to compute the shadows cast by a torus, surpassing conventional methods in both accuracy and ease of use.

The accuracy of the proposed approach was thoroughly assessed by comparing the results with shades generated by SketchUp software, which employs real-time shadow engine[2] for precise shading and shadow casts. The proposed method demonstrated exceptional precision while aligning with the shades produced by SketchUp.

Notably, this proposed approach represents a substantial advancement in the realm of torus shade casting compared to traditional methods. It also holds promise for broader applications, such as the generation of 3D models from architectural shade casts, for instance, from satellite images; thanks to its high level of accuracy. This innovation hope will also empower architects and designers to visualize their creations with greater realism and facilitate effective communication of design concepts. Not only this, the shown process will help them to re-check, re-deign, and revisit existing shade casting methods in an expectation of simplifying the traditional skiagraphy processes.

---

[2] SketchUp's proprietary advanced, real-time Shadow Engine enables precise shade studies of architectural models. By simply inputting the date, time of day, and the model's location on the globe, one can accurately calculate the position of cast shadows. SketchUp simplifies geo-locating an architectural model, giving it a physical location. The Shadow Settings panel handles the rest—by adjusting the Time and Date sliders, one can observe real-time changes in cast shadows.
However, the Shadows feature provides only a general idea of how the sun and shadows will appear at a specific location. The time is not adjusted for daylight saving time, and if the model is geolocated in an area where time zone lines are irregular, the time zone may be off by an hour or more. Since no documentation mentions the exact algorithm, the shadow output suggests a possible use of ray tracing techniques from the sun to calculate shadows. In essence, studying shadows in SketchUp functions similarly to using a virtual heliodon.
Sources [accessed, July 2024]:
https://forums.sketchup.com/t/shadow-analysis-possible-with-this-free-software/11767
https://help.sketchup.com/en/sketchup/casting-real-world-shadows

Such effort will help uphold the artistic value of the traditional shade-casting process in the generations to come.

# 5. Conflict of Interest

The author declares no conflict of interest.